\documentclass[preprint,preprintnumbers,nofootinbib]{revtex4}

\usepackage[dvips]{color}
\usepackage{graphicx}
\usepackage{bbm}

\begin{document}

\draft
\def\thefootnote{\fnsymbol{footnote}}

\begin{center}
{\large\bf Intrinsic Deviation from the Tri-bimaximal Neutrino
Mixing \\ in a Class of $A^{}_4$ Flavor Models}
\end{center}

\begin{center}
{\bf Takeshi Araki}$^{\ a)}$
\footnote{E-mail: araki@ihep.ac.cn},
~{\bf Jianwei Mei}$^{\ b)}$, ~{\bf Zhi-zhong Xing}$^{\ a)}$
\footnote{E-mail: xingzz@ihep.ac.cn}
\end{center}

\begin{center}
{$^{a)}$Institute of High Energy Physics, Chinese Academy of Sciences,
Beijing 100049, China \\
$^{b)}$Max-Planck-Institut f$\rm\ddot{u}$r Gravitationsphysik
(Albert-Einstein-Institut), Golm, Germany}
\end{center}

\begin{abstract}
It is well known that the tri-bimaximal neutrino mixing pattern
$V^{}_0$ can be derived from a class of flavor models with the
non-Abelian $A^{}_4$ symmetry. We point out that small corrections
to $V^{}_0$, which are inherent in the $A^{}_4$ models and arise
from both the charged-lepton and neutrino sectors, have been omitted
in the previous works. We show that such corrections may lead the
$3\times 3$ neutrino mixing matrix $V$ to a non-unitary deviation
from $V^{}_0$, but they cannot result in a nonzero value of
$\theta^{}_{13}$ or any new CP-violating phases. Current
experimental constraints on the unitarity of $V$ allow us to
constrain the model parameters to some extent.
\begin{center}
PACS number(s): 14.60.Pq, 13.10.+q, 25.30.Pt
\end{center}
\end{abstract}

\maketitle

\def\thefootnote{\arabic{footnote}}
\setcounter{footnote}{0}

\section{Introduction}

Thanks to a number of convincing neutrino oscillation experiments
\cite{PDG10}, we have known two neutrino mass-squared differences
($\Delta m^2_{21}$ and $|\Delta m^2_{31}|$) and two neutrino mixing
angles ($\theta^{}_{12}$ and $\theta^{}_{23}$) to a good degree of
accuracy \cite{GG}. The smallest neutrino mixing angle
$\theta^{}_{13}$ remains unknown, but there are some preliminary
hints that it might not be very small (e.g., $\theta^{}_{13} \sim
7^\circ$ \cite{GG,Fogli,KM}). Nevertheless, current experimental
data are consistent very well with a constant neutrino mixing matrix
--- the so-called tri-bimaximal mixing pattern \cite{TB}
\begin{eqnarray}
V^{}_0 = U^T_\omega U^*_\nu = \frac{1}{\sqrt{6}} \ Q^{}_l
\left( \matrix{2 & \sqrt{2} & 0 \cr -1 & \sqrt{2} & \sqrt{3} \cr
1 & -\sqrt{2} & \sqrt{3} \cr} \right) Q^{}_\nu \; ,
\end{eqnarray}
where
\begin{eqnarray}
U^{}_{\omega} & = & \frac{1}{\sqrt{3}}
\left(\begin{array}{ccc}
1 & 1 & 1 \\
1 & \omega & \omega^2 \\
1 & ~\omega^2~ & \omega
\end{array}\right) \; ,
\nonumber \\
U^{}_\nu & = & \frac{1}{\sqrt{2}}
\left(\begin{array}{ccc}
1 & 0 & -1 \\
0 & \sqrt{2} & 0 \\
1 & 0 & 1
\end{array}\right) \; ,
\end{eqnarray}
$\omega = e^{i 2\pi/3}$, $Q^{}_l = {\rm Diag}\{ 1, \omega, -\omega^2
\}$ and $Q^{}_\nu = {\rm Diag}\{1, 1, i\}$ \cite{Xing08}. The
diagonal phase matrix $Q^{}_l$ can be rotated away by redefining the
phases of three charged-lepton fields, but $Q^{}_\nu$ may affect the
neutrinoless double-beta decay if neutrinos are the Majorana
particles. Given the standard parametrization of the
Maki-Nakagawa-Sakata-Pontecorvo (MNSP) neutrino mixing matrix
\cite{MNSP}, $V^{}_0$ corresponds to $\theta^{}_{12} =
\arctan(1/\sqrt{2}) \approx 35.3^\circ$, $\theta^{}_{13} = 0^\circ$
and $\theta^{}_{23} = 45^\circ$. A more realistic form of the MNSP
matrix $V$ is expected to slightly deviate from $V^{}_0$ due to some
nontrivial perturbations\footnote{
For instance, a possible interrelation with the
quark-lepton complementarity is discussed in 
Ref. \cite{ryo}.
}, such that both nonzero $\theta^{}_{13}$
and CP violation can emerge.

It is possible to derive the tri-bimaximal mixing pattern $V^{}_0$
from some neutrino mass models with certain flavor symmetries
\cite{Symmetry}. In this connection the earliest and most popular
application is the non-Abelian discrete $A^{}_4$ symmetry (see,
e.g., Refs. \cite{Ma,Altarelli,Babu}). But the neutrino mixing
matrix derived from a specific $A^{}_4$ model is in general not
equal to $V^{}_0$ unless some approximations are made. In other
words, small corrections to $V^{}_0$ are generally inherent in the
$A^{}_4$ models and can arise both from the charged-lepton sector
and from the neutrino sector. This observation is particularly
interesting for an $A^{}_4$ model built in the vicinity of the TeV
scale, because the resultant corrections to $V^{}_0$ may not be
strongly suppressed. We show that such corrections can lead the
$3\times 3$ neutrino mixing matrix $V$ to a non-unitary deviation
from $V^{}_0$, although they cannot give rise to a nonzero value of
$\theta^{}_{13}$ or any new CP-violating phases. We find that
current experimental constraints on the unitarity of $V$ allow us to
constrain the parameters of an $A^{}_4$ model to some extent.

The remaining part of this paper is organized as follows. In section
II we first outline the salient features of a typical $A^{}_4$ model
and then diagonalize the $6\times 6$ mass matrices of charged
leptons and neutrinos. We show that both $U^{}_\omega$ and
$U^{}_\nu$ in Eq. (2) get modified in this framework. In section III
we work out the non-unitary departure of the resultant $3\times 3$
MNSP matrix $V$ from the tri-bimaximal mixing pattern $V^{}_0 =
U^T_\omega U^*_\nu$. We also constrain the model parameters to some
extent by taking account of current experimental constraints on the
unitarity of $V$. Section IV is devoted to a summary and some
concluding remarks.

\section{Corrections to $U^{}_\omega$ and $U^{}_\nu$ in a typical
$A^{}_4$ model}

Let us consider a simple but typical $A^{}_4$ model proposed by Babu
and He in Ref. \cite{Babu}. The model is an extension of the
standard electroweak $SU(2)^{}_{\rm L} \times U(1)^{}_{\rm Y}$ model
with some additional particles, and it is supersymmetric and $A^{}_4
\times Z^{}_4 \times Z^{}_3$-invariant. The particle content and
charge assignments are summarized in Table I. The discrete
symmetries force the superpotentials of quarks and leptons to have
the following forms:
\begin{eqnarray}
W^{}_q & = & y^d_{ij} Q^{}_i d^c_j H^{}_d +
y^u_{ij} Q^{}_i u^c_j H^{}_u \; ,
\nonumber \\
W^{}_\ell & = & M^{}_E E^{}_i E^c_i + f^{}_\ell L^{}_i E^c_i H^{}_d
+ h^{}_e \left(E^{}_1 \chi^{}_1 + E^{}_2 \chi^{}_2 +
E^{}_3 \chi^{}_3 \right) e^c_1
\nonumber \\
&& + h^{}_\mu \left(E^{}_1 \chi^{}_1 + \omega E^{}_2 \chi^{}_2
+ \omega^2 E^{}_3 \chi^{}_3 \right) e^c_2
+ h^{}_\tau \left(E^{}_1 \chi^{}_1 + \omega^2 E^{}_2 \chi^{}_2
+ \omega E^{}_3 \chi^{}_3 \right) e^c_3 \; ,
\nonumber \\
W^{}_\nu & = & f^{}_\nu L^{}_i \nu^c_i H^{}_u
+ \frac{1}{2} f^{}_{S^{}_a} \nu^c_i \nu^c_i S^{}_a
+ \frac{1}{2}f^{}_{S^{}_b} \nu^c_i \nu^c_i S^{}_b
\nonumber \\
&& + \frac{1}{2}f^{}_{\chi^\prime} \left[
\left(\nu^c_2 \nu^c_3 + \nu^c_3 \nu^c_2 \right)
\chi^\prime_1
+ \left(\nu^c_1 \nu^c_3 + \nu^c_3 \nu^c_1 \right)
\chi^\prime_2
+ \left(\nu^c_2 \nu^c_1 + \nu^c_1 \nu^c_2 \right)
\chi^\prime_3
\right] \; ,
\end{eqnarray}
where the notations are self-explanatory \cite{Babu}.
Note that the quark sector is completely the same as that in the
minimal supersymmetric standard model,
and the $Z^{}_4$ symmetry works as an R-parity such that the
superpotentials possess two units of charge.
Thanks to the supersymmetry and new scalars in Eq. (3), it is
possible to obtain the vacuum expectation values \cite{Babu}
\begin{eqnarray}
&& \langle S^{}_a \rangle = 0 \; , ~~~
\langle S^{}_b \rangle = v^{}_s \; , ~~~
\langle H^{}_u \rangle = v^{}_u \; , ~~~
\langle H^{}_d \rangle = v^{}_d \; , ~~~
\nonumber \\
&& \langle \chi \rangle = \left(v^{}_\chi , \ v^{}_\chi ,
\ v^{}_\chi \right) \; , ~~~
\langle \chi^\prime \rangle =
\left(0 , \ v^{}_{\chi^\prime} , \ 0 \right) \; ,
\end{eqnarray}
where $v^2_u + v^2_d = v^2$ with $v \simeq 174$ GeV.
Thus the $A^{}_4$ symmetry is broken after $\chi$ and
$\chi^\prime$ develop their vacuum expectation values.
\begin{table}[t]
\begin{center}
\caption{The particle content and charge assignments of the model
\cite{Babu}, where the subscript $i$ (for $i=1,2,3$) stands for the
family index.} \vspace{0.5cm}
\begin{tabular}{|c||c|c|c|c|c|c|c|c|c|c|c|c|c|}\hline
           & $Q^{}_i$ & $d^c_i$ & $u^c_i$ & $L^{}_i$ &
             $e^c_1,\ e^c_2,\ e^c_3$ & $\nu^c_i$ &
             $E^{}_i$ & $E^c_i$ & $H^{}_u$ & $H^{}_d$ &
             $\chi^{}_i$ & $\chi^\prime_i$ & $S^{}_{a,b}$ \\ \hline
 $SU(2)^{}_{\rm L}$ & $2$ &  $1$ & $1$ & $2$ & $1$ & $1$ &
             $1$ & $1$ & $2$ & $2$ &
             $1$ & $1$ & $1$ \\ \hline
 $U(1)^{}_{\rm Y}$  & $1/3$ &  $2/3$ & $-4/3$ & $-1$ & $2$ & $0$ &
             $-2$ & $2$ & $1$ & $-1$ &
             $0$ & $0$ & $0$ \\ \hline
 $A^{}_4$     & $1$ & $1$ & $1$ & $3$ & $1,1^{'},1^{''}$ & $3$ &
             $3$ & $3$ & $1$ & $1$ &
             $3$ & $3$ & $1$ \\ \hline
 $Z^{}_4$     & $1$ & $1$ & $0$ & $1$ & $3$ & $0$ &
             $1$ & $1$ & $1$ & $0$ &
             $2$ & $2$ & $2$ \\ \hline
 $Z^{}_3$     & $1$ & $2$ & $0$ & $0$ & $0$ & $1$ &
             $0$ & $0$ & $2$ & $0$ &
             $0$ & $1$ & $1$ \\ \hline
\end{tabular}
\end{center}
\end{table}

In the basis of $(e,E)$ versus $(e^c, E^c)^T$, we obtain the
$6\times 6$ mass matrix of charged leptons from Eqs. (3) and (4):
\begin{eqnarray}
{\cal M}^{}_{\ell E} = \left(\begin{array}{cc}
{\bf 0} & ~f^{}_\ell v^{}_d {\bf 1} \\
H & ~M^{}_E {\bf 1}
\end{array}\right) \; ,
\end{eqnarray}
where ${\bf 1}$ denotes the $3\times 3$ identity matrix, and
\begin{eqnarray}
H = \left(\begin{array}{ccc}
h^{}_e & h^{}_\mu & h^{}_\tau \\
h^{}_e & \omega h^{}_\mu & \omega^2 h^{}_\tau \\
h^{}_e & \omega^2 h^{}_\mu & \omega h^{}_\tau
\end{array}\right) v^{}_\chi
= \sqrt{3} \ U^{}_{\omega}
\left(\begin{array}{ccc}
h^{}_e & 0 & 0 \\
0 & h^{}_\mu & 0 \\
0 & 0 & h^{}_\tau
\end{array}\right) v^{}_\chi \; .
\end{eqnarray}
Note that $f^{}_\ell$, $M^{}_E$ and $h^{}_{\alpha}$ (for $\alpha =
e, \mu, \tau$) can all be arranged to be real in a suitable phase
convention, and the mass scale $M^{}_E$ is assumed to be extremely
large in comparison with the magnitudes of $f^{}_\ell v^{}_d$ and
$h^{}_\alpha v^{}_\chi$. The $6\times 6$ Hermitian matrix ${\cal
M}^{}_{\ell E} {\cal M}_{\ell E}^\dagger$ can be diagonalized via
the unitary transformation $V_l^\dagger {\cal M}^{}_{\ell E} {\cal
M}_{\ell E}^\dagger V^{}_l$, where $V^{}_l$ is given by
\begin{eqnarray}
V^{}_l \simeq \left(\begin{array}{cc} {\bf 1} +
\displaystyle\frac{HH^\dagger}{M_E^2} &
\displaystyle\frac{f^{}_\ell v^{}_d}{M_E} {\bf 1} \\
-\displaystyle\frac{f^{}_\ell v^{}_d}{M_E} {\bf 1} &
{\bf 1} + \displaystyle\frac{HH^\dagger}{M_E^2}
\end{array}\right)
\left(\begin{array}{cc}
U^{}_\omega & {\bf 0} \\
{\bf 0} & {\bf 1}
\end{array}\right) \;
\end{eqnarray}
as a good approximation. The masses of three standard charged
leptons turn out to be
\begin{eqnarray}
m^{}_\alpha \simeq \sqrt{3}\ \frac{f^{}_\ell v^{}_d}{M^{}_E} \
v^{}_\chi h^{}_\alpha \; ,
\end{eqnarray}
where $\alpha$ runs over $e$, $\mu$ and $\tau$. Eq. (7) shows that
$U^{}_\omega$ receives a small correction:
\begin{eqnarray}
U^{}_\omega \longrightarrow U^\prime_\omega = \left( {\bf 1} +
\displaystyle\frac{HH^\dagger}{M_E^2} \right) U^{}_\omega \; .
\end{eqnarray}
It is actually $U^\prime_\omega$ that characterizes the contribution
of charged leptons to the lepton flavor mixing in this $A^{}_4$
model.

Now we turn to the neutrino sector. The type-I seesaw mechanism
\cite{seesaw} is implemented in the $A^{}_4$ model under
consideration, and thus the overall neutrino mass matrix is a
symmetric $6\times 6$ matrix:
\begin{eqnarray}
{\cal M}^{}_{\nu\nu^c} = \left(\begin{array}{cc}
{\bf 0} & f^{}_\nu v^{}_u {\bf 1} \\
f^{}_\nu v^{}_u {\bf 1} & M^{}_R
\end{array}\right) \; ,
\end{eqnarray}
where $M^{}_R$ takes the form
\begin{eqnarray}
M^{}_R = \left(\begin{array}{ccc}
f^{}_{S^{}_b} v^{}_s & 0 & f^{}_{\chi^\prime} v^{}_{\chi^\prime} \\
0 & f^{}_{S^{}_b} v^{}_s & 0 \\
f^{}_{\chi^\prime} v^{}_{\chi^\prime} & 0 & f^{}_{S^{}_b} v^{}_s
\end{array}\right) \; .
\end{eqnarray}
The symmetric neutrino mass matrix in Eq. (10) can be diagonalized
via the orthogonal transformation $V_\nu^T {\cal
M}^{}_{\nu\nu^c}V^{}_\nu$, where the unitary matrix $V^{}_\nu$ is
given by
\begin{eqnarray}
V^{}_\nu \simeq \left(\begin{array}{cc} {\bf 1} - \displaystyle
\frac{1}{2}\cdot\frac{|f^{}_\nu|^2 v_u^2}{M_R^* M_R^T} &
\displaystyle\frac{f^{*}_\nu v^{}_u}{M_R^*} \\
-\displaystyle\frac{f^{}_\nu v^{}_u}{M^{}_R} & {\bf 1} -
\displaystyle\frac{1}{2}\cdot\frac{|f^{}_\nu|^2 v_u^2}{M_R^T M_R^*}
\end{array}\right)
\left(\begin{array}{cc}
U^{}_\nu P^{}_\nu & 0 \\
0 & U^{}_R
\end{array}\right) \;
\end{eqnarray}
to a good degree of accuracy. In this expression $U^{}_\nu$ has
been given in Eq. (2), $P^{}_\nu$ denotes a diagonal phase matrix
\cite{Babu}, and $U^{}_R$ is a unitary matrix responsible for
the diagonalization of $M^{}_R$. The masses of three light (active)
neutrinos turn out to be
$m^{}_1 \simeq \left|m^{}_0 \left(1 + x \right)\right|$,
$m^{}_2 \simeq \left|m^{}_0 \left(1 + x \right)
\left( 1 - x \right)\right|$ and
$m^{}_3 \simeq \left|m^{}_0 \left(1 - x\right)\right|$,
where
\begin{eqnarray}
m^{}_0 = \frac{f_\nu^2 v_u^2 f^{}_{S^{}_b} v^{}_s}
{f_{S^{}_b}^2 v_s^2 - f_{\chi^\prime}^2 v_{\chi^\prime}^2} \; , ~~~
x = -\frac{f^{}_{\chi^\prime} v^{}_{\chi^\prime}}
{f^{}_{S^{}_b} v^{}_s} \; .
\end{eqnarray}
Because both $m^{}_0$ and $x$ are complex, it is possible to adjust
their magnitudes and phases such that the resultant values of
$m^{}_i$ (for $i=1,2,3$) satisfy current experimental data on the
neutrino mass spectrum \cite{Babu}. Eq. (12) shows that $U^{}_\nu
P^{}_\nu$, which signifies the contribution of neutrinos to the
lepton flavor mixing, receives a small correction:
\begin{eqnarray}
U^{}_\nu P^{}_\nu \longrightarrow U^\prime_\nu P^{}_\nu =
\left( {\bf 1} - \displaystyle
\frac{1}{2}\cdot\frac{|f^{}_\nu|^2 v_u^2}{M_R^* M_R^T} \right)
U^{}_\nu P^{}_\nu \; .
\end{eqnarray}
In other words, $U^\prime_\nu$ is not exactly unitary and its
departure from $U^{}_\nu$ is in general an unavoidable consequence
in the type-I seesaw mechanism \cite{Xing09}.

\section{Non-unitary corrections to $V^{}_0$}

With the help of the results obtained in Eqs. (9) and (14), we are
able to calculate the MNSP matrix $V = {U^\prime_\omega}^T
\left(U^\prime_\nu P^{}_\nu\right)^*$ and demonstrate its
non-unitary deviation from the tri-bimaximal mixing pattern
$V^{}_0$. We find
\begin{eqnarray}
V &=& U_\omega^T \left({\bf 1} + \frac{H^{*}H^T}{M_E^2}\right)
\left( {\bf 1} - \frac{1}{2}\cdot \frac{|f^{}_\nu|^2 v_u^2}
{M^{}_R M_R^\dagger} \right) U_\nu^{*} P^*_\nu
\nonumber \\
&\simeq & V^{}_0 P^*_\nu
+ \frac{1}{f^2_\ell v^2_d}
   \left(\begin{array}{ccc}
   m_e^2 & 0 & 0 \\
   0 & m_\mu^2 & 0 \\
   0 & 0 & m_\tau^2
   \end{array}\right) V^{}_0 P^*_\nu
- \frac{1}{2}\cdot \frac{1}{|f^{}_\nu|^2 v_u^2} \ V^{}_0
   \left(\begin{array}{ccc}
   m_1^2 & 0 & 0 \\
   0 & m_2^2 & 0 \\
   0 & 0 & m_3^2
   \end{array}\right) P^*_\nu
\nonumber \\
& \simeq & Q^{}_l \left[ {\bf 1} + \frac{1}{f^2_\ell v^2_d}
   \left(\begin{array}{ccc}
   m_e^2 & 0 & 0 \\
   0 & m_\mu^2 & 0 \\
   0 & 0 & m_\tau^2
   \end{array}\right)
\right. \nonumber \\
&&
\left.
- \frac{1}{12}\cdot \frac{1}{|f^{}_\nu|^2 v_u^2}
   \left(\begin{array}{ccc}
   2\left(2 m_1^2 + m_2^2\right) & 2\left(m_2^2 - m_1^2\right)
   & 2\left(m_1^2 - m_2^2\right) \\
   2\left(m_2^2 - m_1^2\right) & m_1^2 + 2m_2^2 + 3m_3^2 &
                      3m^2_3 -m_1^2 - 2m_2^2 \\
    2\left(m_1^2 - m_2^2\right) & ~ 3m^2_3 -m_1^2 - 2m_2^2 ~ &
                      m_1^2 + 2m_2^2 + 3m_3^2
   \end{array}\right)
\right] V^\prime_0 P^*_\nu \; , ~~~~~
\end{eqnarray}
where
\begin{eqnarray}
V^\prime_0 = Q^*_l V^{}_0 = \frac{1}{\sqrt{6}} \
\left( \matrix{2 & \sqrt{2} & 0 \cr -1 & \sqrt{2} & \sqrt{3} \cr
1 & -\sqrt{2} & \sqrt{3} \cr} \right) Q^{}_\nu \; ,
\end{eqnarray}
and $Q^{}_l$ and $Q^{}_\nu$ have been given below Eq. (2). In
obtaining Eq. (15) we have omitted the higher-order and much smaller
corrections. Because of $v^{}_u = v \sin\beta$ and $v^{}_d = v
\cos\beta$ in the supersymmetric $A^{}_4$ model under consideration,
$v^{}_d \ll v^{}_u$ might hold for a very large value of
$\tan\beta$. Depending on the magnitudes of $f^2_\ell$ and
$|f^{}_\nu|^2$, the term proportional to $1/(f^2_\ell v^2_d)$ or
$1/(|f^{}_\nu|^2 v^2_u)$ in Eq. (15) might not be negligibly small.
These two terms, which are inherent in the model itself, measure the
non-unitary contribution to $V$ or the departure of $V$ from
$V^\prime_0 P^*_\nu$. This observation makes sense since it
indicates that the exact tri-bimaximal neutrino mixing pattern
$V^{}_0$ is not an exact consequence of a class of $A^{}_4$ flavor
models.

One may parametrize the analytical result obtained in Eq. (15) as
follows:
\begin{eqnarray}
V = Q^{}_l \left({\bf 1} -\eta\right)V^\prime_0 P^*_\nu = V^{}_0
P^*_\nu - Q^{}_l \eta V^\prime_0 P^*_\nu \; ,
\end{eqnarray}
where the Hermitian matrix $\eta$ signifies the non-unitary
deviation of $V$ from $V^{}_0 P^*_\nu$. Note that the diagonal phase
matrix $Q^{}_l$ in $V$ can always be rotated away through a
redefinition of the phases of three charged leptons, and the
diagonal phase matrices $Q^{}_\nu$ and $P^*_\nu$ in $V$ only provide
us with the Majorana phases which have nothing to do with leptonic
CP violation in neutrino oscillations. Note also that $\eta$ itself
is real in this $A^{}_4$ model, as one can easily see from Eq. (15),
and thus the unitarity violation of $V$ does not give rise to any
new CP-violating phases. Moreover, it is impossible to obtain
nonzero $V^{}_{e3}$ or $\theta^{}_{13}$ from this typical $A^{}_4$
model, simply because $\eta^{}_{e\mu} = -\eta^{}_{e\tau}$ holds.
Such a disappointing observation implies that the residual flavor
symmetry remains powerful to keep $V^{}_{e3}$ or $\theta^{}_{13}$
vanishing and forbid CP violation, even though the MNSP matrix $V$
is not exactly unitary.

Current experimental data allow us to constrain the matrix elements of
$\eta$ and then constrain the model parameters to some extent.
A recent analysis yields \cite{non-uni}
\begin{eqnarray}
|\eta | <
\left(\begin{array}{ccccc}
2.0\times 10^{-3} & & 6.0\times 10^{-5} & & 1.6\times 10^{-3} \\
6.0\times 10^{-5} & & 8.0\times 10^{-4} & & 1.1\times 10^{-3} \\
1.6\times 10^{-3} & & 1.1\times 10^{-3} & & 2.7\times 10^{-3}
\end{array}\right) \; .
\end{eqnarray}
In view of Eqs. (15) and (16), we immediately obtain
\begin{eqnarray}
\eta^{}_{e \mu} & = & -\eta^{}_{e \tau} = \frac{\Delta m^2_{21}} {6
|f^{}_\nu|^2 v^2_u} = \frac{\Delta m^2_{21}} {6 |f^{}_\nu|^2 v^2
\sin^2\beta} \; ,
\nonumber \\
\eta^{}_{\mu \tau} & = & \frac{\Delta m_{31}^2 + 2 \Delta m_{32}^2}
{12|f^{}_\nu|^2 v^2_u} \simeq \frac{\Delta m_{31}^2} {4|f^{}_\nu|^2
v^2 \sin^2\beta} \; ,
\end{eqnarray}
where $\Delta m^2_{21} \equiv m^2_2 - m^2_1 \simeq 7.6 \times
10^{-5} ~{\rm eV}^2$ and $\Delta m^2_{31} \equiv m^2_3 - m^2_1
\simeq m^2_3 - m^2_2 \equiv \Delta m^2_{32} \simeq \pm 2.4 \times
10^{-3} ~{\rm eV}^2$ \cite{GG}. Eq. (19) leads us to a simple but
instructive relation for three off-diagonal matrix elements of
$\eta$:
\begin{eqnarray}
\frac{\eta^{}_{e\mu}}{\eta^{}_{\mu \tau}} =
-\frac{\eta^{}_{e\tau}}{\eta^{}_{\mu \tau}} \simeq \frac{2}{3}
\cdot \frac{\Delta m^2_{21}}{\Delta m^2_{31}} \; .
\end{eqnarray}
Therefore, $|\eta^{}_{e\mu}|/|\eta^{}_{\mu\tau}| =
|\eta^{}_{e\tau}|/|\eta^{}_{\mu\tau}| \simeq 2.1 \times 10^{-2}$.
Comparing this prediction with Eq. (18), one may self-consistently
get $|\eta^{}_{e\mu}| = |\eta^{}_{e\tau}| < 2.3 \times 10^{-5}$
by taking $|\eta^{}_{\mu\tau}| < 1.1 \times 10^{-3}$. So it is
more appropriate to use the upper bound of $|\eta^{}_{\mu\tau}|$
to constrain the lower bound of $|f^{}_\nu|$ by means of Eq. (19).
We arrive at
\begin{eqnarray}
|f^{}_\nu| = \frac{1}{2 v \sin\beta} \cdot\frac{\sqrt{|\Delta
m^2_{31}|}} {\sqrt{|\eta^{}_{\mu \tau}|}} \ > \
\frac{4.2}{\sin\beta} \times 10^{-12} \; .
\end{eqnarray}
This result, which depends on the value of $\tan\beta$ in the
supersymmetric $A^{}_4$ model, implies that the Yukawa coupling of
neutrinos should not be too small in order to preserve the unitarity
of $V$ at an experimentally-allowed level. It clearly indicates that
an arbitrary choice of $f^{}_\nu$ in the neglect of small unitarity
violation of $V$ is inappropriate for model building, because the
correlation between $f^{}_\nu$ and the deviation of $V$ from the
tri-bimaximal mixing pattern is an intrinsic property of a class of
$A^{}_4$ models.

The diagonal matrix elements of $\eta$ consist of the contributions
from both the charged-lepton sector and the neutrino sector, as shown
in Eq. (15). Their competition depends on the sizes of $f^{}_\ell$,
$f^{}_\nu$ and $\tan\beta$. For simplicity, here we assume that the
charged-lepton contribution to $\eta^{}_{\alpha\alpha}$ (for $\alpha
=e, \mu, \tau$) is dominant. Then it is straightforward to obtain
\begin{eqnarray}
\eta^{}_{\alpha \alpha} \simeq -\frac{m^2_\alpha}{f^2_\ell v^2_d} =
-\frac{m^2_\alpha}{f^2_\ell v^2 \cos^2\beta} \; .
\end{eqnarray}
As a result,
\begin{eqnarray}
\eta^{}_{ee} : \eta^{}_{\mu\mu} : \eta^{}_{\tau\tau} \simeq
m^2_e : m^2_\mu : m^2_\tau \simeq 1: 44566: 12880040 \; ,
\end{eqnarray}
where we have input the central values of three charged-lepton masses
at the electroweak scale \cite{XZZ}. Comparing this prediction with
Eq. (18), one may self-consistently arrive at
$|\eta^{}_{ee}| < 2.1 \times 10^{-10}$ and
$|\eta^{}_{\mu\mu}| < 9.3 \times 10^{-6}$ by taking
$|\eta^{}_{\tau\tau}| < 2.7 \times 10^{-3}$. It is therefore
more appropriate to use the upper bound of $|\eta^{}_{\tau\tau}|$
to constrain the lower bound of $|f^{}_\ell|$ with the help of
Eq. (22). We find
\begin{eqnarray}
|f^{}_\ell| \simeq \frac{m^{}_\tau}{v \cos\beta
\sqrt{|\eta^{}_{\tau\tau}|}} > \frac{0.19}{\cos\beta} \; ,
\end{eqnarray}
where $m^{}_\tau \simeq 1746.24$ MeV has been input at the
electroweak scale \cite{XZZ}. This result, which also depends on the
value of $\tan\beta$ in the supersymmetric $A^{}_4$ model, shows
that the Yukawa coupling of charged leptons should be relatively
large in order to preserve the unitarity of $V$ as constrained by
current measurements. We stress that an arbitrary choice of either
$f^{}_\ell$ or $f^{}_\nu$ in the neglect of small unitarity
violation of $V$ might be problematic for model building, simply
because they receive constraints both from the model itself and from
the experimental data. In this sense one must be cautious to claim
that an $A^{}_4$ flavor model can predict the tri-bimaximal neutrino
mixing pattern whose matrix elements are constant and thus have
nothing to do with the model parameters \cite{lam}. In fact, the
slight (non-unitary) deviation of $V$ from the tri-bimaximal mixing
pattern is likely to impose a strong restriction on some model
parameters like $f^{}_\ell$, $f^{}_\nu$ and $\tan\beta$.

\section{Summary}

We have examined a class of $A^{}_4$ flavor models to see whether
the tri-bimaximal neutrino mixing pattern $V^{}_0$ is an exact
consequence of such models. We find that small corrections to
$V^{}_0$ are actually inherent in the $A^{}_4$ models and may arise
from both the charged-lepton and neutrino sectors. We have
demonstrated that such corrections may lead the MNSP matrix $V$ to a
non-unitary deviation from $V^{}_0$, but they cannot result in a
nonzero $V^{}_{e3}$ (or $\theta^{}_{13}$) or any new CP-violating
phases. In particular, the slight unitarity violation of $V$ is
sensitive to several model parameters, including the Yukawa
couplings of charged leptons and neutrinos. We have shown that
current experimental constraints on the unitarity of $V$ allow us to
constrain the model parameters to some extent.

We stress that the departure of $V$ from $V^{}_0$ explored in this
work is an intrinsic property of a class of flavor models with the
non-Abelian $A^{}_4$ symmetry. Different departures may result
either from the vacuum-expectation-value misalignments in a certain
$A^{}_4$ model or from some purely phenomenological perturbations
\cite{Barry}. The non-unitary deviation of $V$ from $V^{}_0$ is in
some sense more interesting because it might give rise to new
CP-violating effects in a variety of long-baseline neutrino
oscillation experiments \cite{Yasuda}. Since a lot of attention has
been paid to how to derive the tri-bimaximal neutrino mixing pattern
$V^{}_0$, the points revealed in our paper should be taken into
account when one attempts to build specific flavor models with
discrete family symmetries.

\vspace{0.6cm}

One of us (J.M.) is grateful to the TPCSF for financial support
during his visiting stay at the IHEP. This work was supported in
part by the National Natural Science Foundation of China under grant
No. 10425522 and No. 10875131.

\end{document}